\documentclass[conference]{IEEEtran}

\ifCLASSINFOpdf
\else
\fi

\ifCLASSINFOpdf
\usepackage[pdftex]{graphicx}
\graphicspath{{./images/}}
\else
\fi

\usepackage{mathtools}
\usepackage{mathrsfs}
\usepackage{amsmath}
\usepackage{bm}
\usepackage[cmintegrals]{newtxmath}

\ifCLASSOPTIONcompsoc
\usepackage[caption=false,font=normalsize,labelfont=sf,textfont=sf]{subfig}
\else
\usepackage[caption=false,font=footnotesize]{subfig}
\fi
\usepackage{cite}
\usepackage{tabularx}
\usepackage{xcolor}
\usepackage{multirow}

\newcommand{\eexpo}[1]{\exp \left(  #1 \right) }

\newcommand{\eabsn}[2]{\left|  #1 \right| ^{#2}}
\newcommand{\eherm}[1]{\left(  #1 \right)^{H}}
\newcommand{\econj}[1]{\left( #1 \right)^{*}}

\newcommand{\eelem}[3]{\left[  #1 \right]^{#2}_{#3} }

\newcommand{\ediag}[1]{\text{diag}\left( #1 \right)}

\newcommand{\enormn}[3]{\left| \left| #1\right| \right| ^{#2}_{#3}}

\newcommand{\ecsizeo}[1]{\in\mathbb{C}^{#1}}

\newcommand{\ecsize}[2]{\in\mathbb{C}^{#1\times#2}}

\newcommand{\egausd}[2]{\mathcal{CN}\left( #1,#2\right) }

\newcommand{\eunifc}[2]{\mathcal{U}\left[  #1,#2 \right]  }

\newcommand{\eexpabstwo}[1]{\mathbb{E} \left\lbrace  \left| #1 \right|^2 \right\rbrace}

\newcommand{\eexp}[1]{\mathbb{E} \left\lbrace #1 \right\rbrace}

\newcommand{\ereal}[1]{\Re \left\lbrace #1 \right\rbrace}

\newcommand{\ecomplexity}[1]{\mathcal{O}\left( #1\right) }

\def\mydate{\leavevmode\hbox{\the\year/\twodigits\month/\twodigits\day}}
\def\twodigits#1{\ifnum#1<10 0\fi\the#1}

\newcommand{\eAddFig}[4]{
	\begin{figure}[!t]
		\centering
		\includegraphics[width=#2\linewidth]{#1}
		\caption{#4}
		\label{#3}
	\end{figure}
}

\begin{document}
\title{Non-Coherent MIMO-OFDM Uplink empowered by the Spatial Diversity in Reflecting Surfaces}
\author{\IEEEauthorblockN{Kun Chen-Hu$^1$, George C. Alexandropoulos$^2$, and Ana García Armada$^1$}
$^1$Department of Signal Theory and Communications, Universidad Carlos III de Madrid, Spain\\
$^2$Department of Informatics and Telecommunications, National and Kapodistrian University of Athens, Greece\\
E-mails: kchen@tsc.uc3m.es, alexandg@di.uoa.gr, agarcia@tsc.uc3m.es}

\maketitle
\begin{abstract}
Reflecting Surfaces (RSs) are being lately envisioned as an energy efficient solution capable of enhancing the signal coverage in cases where obstacles block the direct communication from Base Stations (BSs), especially at high frequency bands due to attenuation loss increase. In the current literature, wireless communications via RSs are exclusively based on traditional coherent demodulation, which necessitates the estimation of accurate Channel State Information (CSI). However, this requirement results in an increased overhead, especially in time-varying channels, which reduces the resources that can be used for data communication. In this paper, we consider the uplink between a single-antenna user and a multi-antenna BS and present a novel RS-empowered Orthogonal Frequency Division Multiplexing (OFDM) communication system based on the differential phase shift keying, which is suitable for high noise and/or mobility scenarios. As a benchmark, analytical expressions for the Signal-to-Interference and Noise Ratio (SINR) of the proposed system are presented. Our extensive simulation results verify the accuracy of the presented analysis and showcase the performance and superiority of the proposed system over coherent demodulation.
\end{abstract}

\IEEEpeerreviewmaketitle

\section{Introduction}
\label{sec:intro}

Reflecting surfaces (RSs) are lightweight and hardware-efficient artificial planar structures of passive reflective elements, capable of reflecting the received signal to several directions. Specifically, Reconfigurable Intelligent Surfaces (RISs) \cite{Ris01}, which are a kind of RSs, are also able to support a wide variety of functionalities \cite{WavePropTCCN}, ranging from controllable absorption, beam and wavefront shaping to polarization control, broadband pulse delay, radio-coverage extension, and harmonic generation. Hence, it is expected to play a significant role in the upcoming 6-th Generation (6G) of mobile communication systems \cite{Samsung}, due to the fact that these surfaces are able to both improve and extend the signal transmitted by either the Base Station (BS) or User Equipment (UE), circumventing the high attenuation loss and signal blockage produced by the use of high frequency bands.

RIS is typically combined with the classical Coherent Demodulation Scheme (CDS) \cite{8879620,RisChanEst04}, where the knowledge of (pefect) Channel State Information (CSI) is required for the optimized configuration of the RIS tunable elements and the demodulation of the signal at the receiving node. However, the benefit of this reconfigurability property comes at the expense of transmitting an enormous amount of signal references to estimate the cascaded channel matrix \cite{8879620}, which encompasses the joint effect of the signal propagation over the BS-RIS and RIS-UE links, due to the fact that the minimum required training periods equals the total number of configurations of the RIS. Therefore, this estimation overhead becomes prohibitive as the numbers of RIS elements and configurations increase \cite{Ris04}. Time Duplex Division (TDD) is typically adopted and the CSI is assumed to be estimated in the uplink and then reused in the downlink. To this end, the coherence time is always considered to be long enough to cope with the channel training and uplink/downlink data transmission stages. Given the CSI availability, the BS computes the best pair of precoder/combiner as well as the set of RIS elements' configuration, which is communicated to the RIS via a side control link. This processing task is not straightforward due to the fact that a non-convex design optimization needs to be solved, increasing the operational complexity of the RIS-empowered communication system. When Orthogonal Frequency Division Multiplexing (OFDM) \cite{ofdm1} is taken into account, the complexity of channel estimation scales with the number of subcarriers \cite{RisChanEst04}. 

The Non-CDS (NCDS) is an alternative demodulation scheme which does not require CSI, and hence, reducing the undesirable signalling overhead and increasing the effective data rate of communication system. This scheme is capable of obtaining the transmitted information by the non-coherent combination of the received signals, no matter the behaviour of the propagation channel. As an additional benefit, it possesses a reduced complexity, which implies cheaper transceiver hardware devices and lower latency for processing. Recently, NCDS has been combined with massive Multiple-Input Multiple-Output (MIMO) systems \cite{Ana2015,Victor2018,Kun2019}, where it was shown to provide a significant performance gain compared to CDS for the challenging scenarios of vehicular and low-latency communications. Motivated by the fact that, communication systems including RISs with large numbers of passive elements require an unreasonable number of reference signals for the cascaded channels estimation. Therefore, we present in this paper a Single-Input Multiple-Output (SIMO) OFDM system with differential Phase Shift Keying (PSK) modulation empowered by a generic RS. Our system does not require channel estimation, and therefore, the overhead produced by the transmission of the reference signals is avoided. Besides, in order to circumvent the non-convex optimization problem for the RIS configuration, we propose the exploitation of the spatial diversity provided by a generic RS, where the surface can be either reconfigurable or not. Hence, the proposed system is not only able to improve the efficiency of the system, but it is also capable of reducing the processing complexity, especially for broadband multi-carrier waveforms, simplifying the massive deployment of reflecting surfaces. We analytically characterize the Signal-to-Interference plus Noise Ratio (SINR) of the proposed NCDS system enhanced by RSs. Our simulation results verify the accuracy of the presented analysis and highlight the superiority of the proposed NCDS system over a relevant CDS one in terms of symbol error probability (SEP). 

\textbf{Notation:} Matrices, vectors, and scalar quantities are denoted by boldface uppercase, boldface lowercase, and normal letters, respectively. $\left[\mathbf{A}\right]_{mn}$ denotes the element in the $m$-th row and $n$-th column of $\mathbf{A}$, $\left[\mathbf{A}\right]_{:,n}$ is $\mathbf{A}$'s $n$-th column, and $\left[\mathbf{a}\right]_{n}$ represents the $n$-th element of $\mathbf{a}$. $\ediag{\mathbf{a}}$ denotes a diagonal matrix whose diagonal elements are formed by $\mathbf{a}$'s elements. $\Re(\cdotp)$ and $\Im(\cdotp)$ represent the real and imaginary part of a complex number, respectively, and $\jmath$ is the imaginary unit, while $*$ denotes the convolution operation. $\enormn{\cdotp}{2}{F}$ denotes the squared Frobenius norm. $\eabsn{\cdotp}{}$ is the absolute value. $\mathbb{E}\left\lbrace \cdotp \right\rbrace$ represents the expected value of a random variable and $\mathcal{CN}(0,\sigma^2)$ represents the circularly-symmetric and zero-mean complex normal distribution with variance $\sigma^2$. $\eunifc{a}{b}$ denotes the continuous uniform distribution with the minimum value $a$ and maximum value $b$.

\section{System Model}\label{sec:system_model}
The mobile communication scenario is built by a BS, an RS, and a single-antenna UE (see Fig. \ref{fig:scenario}). The BS is equipped with a uniform rectangular array (URA) consisting of $B=B_{H}B_{V}$ antenna elements, where $B_{H}$ and $B_{V}$ denote the number of elements in the horizontal and vertical axes, respectively, and the distance between any two contiguous elements in their respective axes is given by $d_{H}^{\text{BS}}$ and $d_{V}^{\text{BS}}$. Analogously to the BS, the RS is built by $M=M_{H}M_{V}$ fully passive reflecting unit elements, whose respective distances between elements are given by $d_{H}^{\text{RS}}$ and $d_{V}^{\text{RS}}$. The UE is constrained to have a single antenna element.

\eAddFig{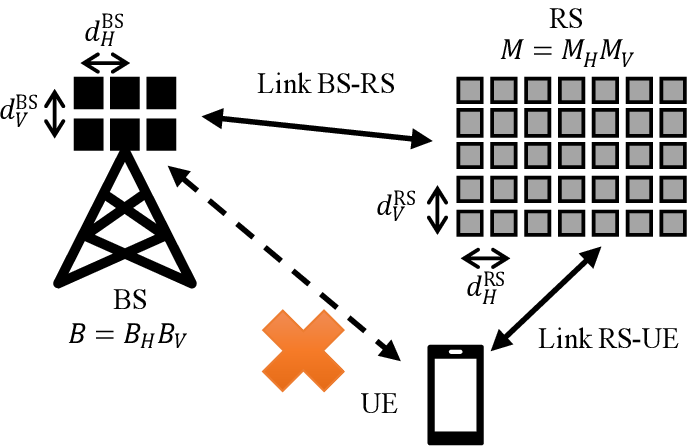}{1}{fig:scenario}{The RS-empowered wireless communication link comprising a multi-antenna BS, a multi-element passive RS, and a single-antenna mobile UE.}

Regarding the signal propagation, the direct communication link between the BS and UE (BS-UE) is assumed to be absent, due to the presence of blockages. Therefore, the information exchanged between the BS and UE should be transmitted through the RS, via the BS-RS and RS-UE communication links. Focusing on the uplink case, the UE transmits both pilot (if needed, e.g., in CDS) and data symbols to the BS through the RS. It is understood that other UEs may be multiplexed in different orthogonal (time or frequency) resources; this extension is left for future work. It is assumed that, at each communication frame, the UE transmits a frame of $N$ contiguous OFDM symbols of $K$ subcarriers each. In order to avoid the Inter-Symbol and Inter-Carrier Interferences (ISI and ICI), the length $L_{CP}$ of the cyclic prefix must be long enough to absorb the effective multipath produced by the cascaded channel, namely the sum of the lengths of each of the channel responses of both BS-RS and RS-UE channels. The baseband representation of the received signal $\mathbf{y}_{k,n}\ecsizeo{B}$ at the BS in the $k$-th subcarrier, with $1\leq k \leq K$, and $n$-th OFDM symbol, with $1 \leq n \leq N$, is given by
\begin{equation}\label{eqn:model_y}
	\mathbf{y}_{k,n} = \mathbf{q}_{k,n}x_{k,n}  + \mathbf{v}_{k,n}, 
\end{equation}
where $x_{k,n}\in\mathbb{C}$ denotes the symbol transmitted from the UE at the $k$-th subcarrier and $n$-th OFDM symbol, whose transmit power is $\eexpabstwo{x}=P_{x}$, $\mathbf{v}_{k,n}\ecsizeo{B}$ represents the Additive White Gaussian Noise (AWGN) vector which is distributed as $\eelem{\mathbf{v}_{k,n}}{}{b}\sim\egausd{0}{\sigma_{v}^{2}}$, and $\mathbf{q}_{k,n}\ecsizeo{B}$ is the effective RS-empowered cascaded channel frequency response, which can be decomposed for $1\leq k \leq K$ and $1 \leq n \leq N$ as
\begin{equation}\label{eqn:model_chan1}
	\mathbf{q}_{k,n}\triangleq\mathbf{H}_{k,n}\mathbf{\Psi}_{n}\mathbf{g}_{k,n}=\sum_{m=1}^{M}\eelem{\boldsymbol{\psi}_{n}}{}{m} \eelem{\mathbf{H}_{k,n}}{}{:,m}\eelem{\mathbf{g}_{k,n}}{}{m},
\end{equation}
where $\mathbf{H}_{k,n}\ecsize{B}{M}$ is the channel frequency response matrix between BS and RS, $\mathbf{g}_{k,n}\ecsizeo{M}$ accounts for the channel frequency response vector between RS and the single UE of interest, and $\mathbf{\Psi}_{n}\triangleq\ediag{\boldsymbol{\psi}_n}\in\mathbb{C}^{M\times M}$ is a diagonal matrix accounting for the effective phase configurations applied by the passive reflecting elements of the RS at the $n$-th OFDM symbol, where $\boldsymbol{\psi}_n\in\mathbb{C}^{M\times M}$ is defined as
\begin{equation}\label{eqn:panel}
	\boldsymbol{\psi}_n \triangleq \begin{bmatrix} \eexpo{\jmath \psi_{n,1}} & \cdots & \eexpo{\jmath\psi_{n,M}} \end{bmatrix}, 
\end{equation}
with $\psi_{n,m}$ for $1\leq m \leq M$ representing the phase shift of the $m$-th passive element of the RS panel. 

The channel frequency response for both links (BS-RS and RS-UE) are independent and identically distributed (IID) random variables. Hence, the propagation channels (BS-RS and RS-UE) at the $k$-th subcarrier and $n$-th OFDM symbol are modelled as
\label{subsec:iid}
\begin{equation}\label{eqn:chan_h_iid}
	\mathbf{H}_{k,n} \triangleq \sqrt{L_{\alpha}} \mathbf{A}_{k,n}, \quad \eelem{\mathbf{A}_{k,n}}{}{bm} \sim \egausd{0}{\sigma_{\alpha}^{2}},
\end{equation}
\begin{equation}\label{eqn:chan_g_iid}
	\mathbf{g}_{k,n} \triangleq \sqrt{L_{\beta}} \mathbf{b}_{k,n}, \quad \eelem{\mathbf{b}_{k,n}}{}{m} \sim \egausd{0}{\sigma_{\beta}^{2}},
\end{equation}
\begin{equation*}
	1\leq b \leq B, \quad 1 \leq m \leq M, 
\end{equation*}
where $L_{\alpha}$ and $L_{\beta}$ denote the large-scale gains of the BS-RS and RS-UE links, respectively, and $\mathbf{A}_{k,n}\ecsize{B}{M}$ and $\mathbf{b}_{k,n}\ecsizeo{M}$ model the small-scale fading for their respective channels, according to a Rayleigh distribution. Hence, the average gain of each link is $\sigma_{h}^{2} = L_{\alpha}\sigma_{\alpha}^{2}$ and $\sigma_{g}^{2} = L_{\beta}\sigma_{\beta}^{2}$, respectively. Moreover, it is assumed that the channel between BS and RS remains quasi-static, while the channel between the RS and UE may suffer from time variability given as
\begin{equation} \label{eq:dopplerchan}
	\eexp{\econj{\eelem{\mathbf{g}_{k,n}}{}{m}}\eelem{\mathbf{g}_{k,n'}}{}{m}}
	=\eabsn{J_{0}\left(2\pi f_{d}  \frac{\Delta n}{\Delta f}\left( 1+\frac{L_{CP}}{K}\right) \right)}{},
\end{equation}
\begin{equation*}
	\Delta n = n'-n, \quad 1 \leq k \leq K, \quad 1 \leq n \leq N, \quad 1 \leq m \leq M,
\end{equation*}
where $J_{0}\left( \cdot \right) $ denotes the zero-th order Bessel function of the first kind \cite{bessel}, and $f_{d}$ and $\Delta f$ represent the Doppler frequency shift experienced by the signal transmitted from the UE and the distance between two contiguous subcarriers, respectively, both measured in Hz.

\section{Proposed RS-Empowered System Based on Differential Modulation}
\label{sec:nc-diff}
The proposed NCDS based on differential modulation \cite{Ana2015,Victor2018,Kun2019} enhanced by an RS not only does not require training-based channel estimation in order to perform the demodulation and decision, but it also avoids the complex non-convex optimization problem to get the phase configurations for the RS. In this section, we detail the steps for performing the modulation and demodulation steps.

At the UE, the data symbols are differentially encoded in the time domain before their transmission as:
\begin{equation} \label{eqn:diff_enc}
	x_{k,n}  =
	\left\{\begin{array}{@{}cl}
		s_{k,n},  & n=1 \\
		x_{k,n-1} s_{k,n}, & 2\leq n \leq N\\
	\end{array}
	\right., \quad 1 \leq k \leq K.
\end{equation}
where $s_{k,n}$ denotes the complex symbol to be transmitted at the $k$-th subcarrier and $n$-th OFDM symbol, that belongs to a PSK constellation and its power is normalized (i.e., $\eabsn{s_{k,n}}{2}=1$). Note that the differential modulation only requires a single reference symbol $s_{k,1}$ at the beginning of the burst in order to allow the differential demodulation, which represents a negligible overhead. The differential modulation can be also implemented in the frequency domain with the same performance \cite{Kun2019}. Before data transmission, the power of differential symbols $x_{k}^{n}$ is scaled according to $P_{x}$.

Given (\ref{eqn:model_y}), the BS performs the differential decoding as
\begin{equation} \label{eqn:diff_dec}
	z_{k,n} = \frac{1}{MB} \eherm{\mathbf{y}_{k,n-1}}\mathbf{y}_{k,n}= \frac{1}{MB}\sum_{i=1}^{4} I_{i}, 
\end{equation}
\begin{equation*}
	2 \leq n \leq N, \quad 1 \leq k \leq K,
\end{equation*}
\begin{equation} \label{eqn:diff_dec1}
	I_{1} = \eherm{\mathbf{q}_{k,n-1}}\mathbf{q}_{k,n} s_{k,n}, \quad I_{2} = \eherm{\mathbf{q}_{n-1}^{k}x_{k,n-1}}\mathbf{v}_{k,n},
\end{equation}
\begin{equation} \label{eqn:diff_dec3}
	I_{3} = \eherm{\mathbf{v}_{k,n-1}}\mathbf{q}_{k,n} x_{k,n}, \quad I_{4} = \eherm{\mathbf{v}_{k,n-1}} \mathbf{v}_{k,n},
\end{equation}
where $I_{1}$ includes the useful symbol $s_{k,n}$ to be decided, however, it is polluted by the effective RS-empowered cascaded channel. In addition, $I_{2}$ and $I_{3}$ represent the cross-interference terms produced by the noise and the received differential symbol in two time instants, while $I_{4}$ is exclusively produced by the product of the noise in two instants. 

The symbol decision is performed over the variable $z_{k,n}$ in \eqref{eqn:diff_dec} without the knowledge of the CSI, and hence, the undesirable cascaded channel estimation can be avoided. Note that our proposal is flexible since it can be used for any kind of RS, either reconfigurable or not. Hence, $\mathbf{\Psi_{n}}$ can be either randomly set without any restriction when the panel is intelligent (RIS) ($\psi_{n,m} \sim \eunifc{0}{2\pi}$), or take advantage the existing configuration set by a non-reconfigurable one. It will be shown in the performance evaluation results that the proposed NCDS-based approach provides substantial gains over the baseline CDS, both in terms of computational complexity and achievable performance.

\section{Analysis of the SINR and complexity}
\label{sec:analysis}
In this section we provide the analysis or the SINR and the complexity comparison with the traditional CDS.

\subsection{Analysis of the SINR}
According to (\ref{eqn:diff_dec})-(\ref{eqn:diff_dec3}), there are interference and noise terms produced by the differential decoding. The received symbol $z_{k,n}$ should be compared to the transmitted symbol $s_{k,n}$ taking into account the channel gain in order to characterize these undesirable effects, which can be expressed as
\begin{equation} \label{eqn:interference}
\begin{split}
& \eexpabstwo{ \sigma_{h}^{2}\sigma_{g}^{2} s_{k,n}-z_{k,n}}= \sigma_{h}^{4}\sigma_{g}^{4} P_{x}^{2} +  \eexpabstwo{z_{k,n}} -\\
& -2\sigma_{h}^{2}\sigma_{g}^{2}\ereal{\eexp{\eherm{s_{k,n}}z_{k,n}}},
\end{split}
\end{equation}
where the expectation is performed over the subcarriers and OFDM symbols. According to \cite{Victor2018}, the four terms given in (\ref{eqn:diff_dec1}) and (\ref{eqn:diff_dec3}) are statistically independent due to the fact that the channel frequency response, noise, and symbols are independent random variables, and the noise samples between two time instants are also independent. Hence, the two terms in (\ref{eqn:interference}) can be simplified as
\begin{equation} \label{eqn:victor1}
\eexpabstwo{z_{k,n}} = \frac{1}{M^2B^2}\sum_{i=1}^{4}\eexpabstwo{I_{i}},
\end{equation}
\begin{equation} \label{eqn:victor2}
\eexp{\eherm{s_{k,n}}z_{k,n}} = \frac{1}{MB}\eexp{\eherm{s_{k,n}}I_{1}},
\end{equation}
and the SINR of the proposed NCDS approach ($\rho_{nc}$) can be defined as
\begin{equation} \label{eqn:sinr}
\begin{split}
\frac{\sigma_{h}^{4}\sigma_{g}^{4}P_{x}^{2}}{\rho_{nc}}=& \sigma_{h}^{4}\sigma_{g}^{4}P_{x}^{2}+\frac{1}{M^{2}B^{2}}\sum_{i=1}^{4}\eexpabstwo{I_{i}} -\\
&-\frac{2}{MB}\sigma_{h}^{2}\sigma_{g}^{2}\ereal{\eexp{\eherm{s_{k,n}}I_{1}}}.
\end{split}
\end{equation}

Assuming the IID Rayleigh channel model and following the manipulations given by \cite{Ana2015,Victor2018,Kun2019}, each of the expected values in (\ref{eqn:sinr}) can be expressed as
\begin{equation} \label{eqn:tlast_iid}
	\eexp{\eherm{s_{k,n}}I_{1}} = P_{x}^{2} B\sigma_{h}^{2} M\sigma_{g}^{2}, \quad \eexpabstwo{I_{4}} = B\sigma_{v}^{4},
\end{equation}
\begin{equation} \label{eqn:t1_iid}
	\eexpabstwo{I_{1}} = P_{x}^{2} (1+B)B\sigma_{h}^{4} (1+M)M\sigma_{g}^{4},
\end{equation}
\begin{equation} \label{eqn:t23_iid}
	\eexpabstwo{I_{2}} = \eexpabstwo{I_{3}} = \sigma_{v}^{2} P_{x} B\sigma_{h}^{2} M\sigma_{g}^{2}.
\end{equation}
Substituting (\ref{eqn:tlast_iid})-(\ref{eqn:t23_iid}) into (\ref{eqn:sinr}), yields the following expression for the SINR of the proposed NCDS approach:
\begin{equation} \label{eqn:diff_snr_iid}
\rho_{nc} =\frac{MB}{B+M+1  + \frac{2\sigma_{v}^{2}}{\sigma_{h}^{2}\sigma_{g}^{2}P_{x}} + \frac{\sigma_{v}^{4}}{\sigma_{h}^{4}\sigma_{g}^{4}P_{x}^{2}M}},
\end{equation}
which indicates that not only the number $B$ of the BS antennas improves the system performance, but also the number $M$ of the RS passive elements helps to reduce the interference and noise terms. Hence, the spatial diversity can be obtained from both the BS and RS.

\begin{table}[]
	\centering
	\caption{Complexity Comparison between the Proposed NCDS and the Considered Baseline CDS.}
	\label{tab:complexity}
	\vspace{-3mm}
	\begin{tabular}{|c|c|c|}
		\hline
		& \textbf{\begin{tabular}[c]{@{}c@{}}System Optimization\\ Complexity\end{tabular}} & \textbf{\begin{tabular}[c]{@{}c@{}}Total Number of \\ Complex Products\end{tabular}} \\ \hline
		\textbf{CDS} \cite{Ris06} & $\ecomplexity{R_{t}(B^{3}+M)K}$                                                     & $BK$                                                                           \\ \hline
		\textbf{NCDS} & $-$                                                                                & $(B+1)(K-1)$                                                                   \\ \hline
	\end{tabular}
\end{table}

\subsection{Complexity Analysis}

The complexity evaluation for both CDS and NCDS are summarized in Table \ref{tab:complexity}. For the NCDS, the differential encoding, the transmitter (i.e., the UE) requires $K-1$ complex products at each OFDM symbol, where the receiver (i.e., BS) needs the same number of complex products for the differential decoding at each RF chain before the symbol decision, resulting in the total of $B(K-1)$ products. On the contrary, the CDS not only requires complex products for performing the post-coding, but it also requires to solve the optimization problem to obtain the BS combining vector and the RS phase configuration. In order to constrain the complexity, the considered baseline CDS implements the iterative method of \cite{Ris06} at the expense of reducing the performance, whose complexity linearly scales with the number of algorithmic iterations required iterations, $R_{t}$, the number $M$ of RS passive elements, and the number $K$ of OFDM subcarriers.

\section{Performance Evaluation Results}
\label{sec:num_res}

Several numerical results are provided in this section in order to show the performance of the proposed NCDS, as compared to the considered baseline CDS, and the accuracy of the analytical results. Additionally, in order to provide a better understanding and show the validity of our proposal, we also evaluate it using a geometric wideband channel model \cite{chan_meas1,chan_meas2}.

\subsection{Baseline RIS-Empowered System Based on CDS}
\label{sec:baseline}
According to \cite{RisChanEst01,RisChanEst02,8879620,RisChanEst04}, RIS-empowered systems based on CDS require to perform a channel training stage before the data transmission stage. This channel training stage mainly consists of three tasks: cascaded channel sounding, optimization for the desired RIS configuration, and parameter feedback. Moreover, previous works \cite{Ris01,Ris02,Ris03,Ris04,Ris05} always assumed that the coherence time ($T_{c}$) is always long enough so that the duration of channel training does not penalize the duration of the data transmission stage. However, even with low mobility, the cascaded channel will suffer from a certain time variability and its estimation must be periodically updated. Consequently, the precoder/combiner and RIS phase configuration need to be updated accordingly. 

To take into account the inefficiency produced by the channel training stage, it is assumed that the data rate penalty due to channel sounding is lower-bounded by taking only into account the channel sounding time, since the feedback time is typically assumed to be negligible, and the optimization time is difficult to quantify due to the fact that the time required for solving the design optimization problem depends on the chosen numerical method and the amount of resources assigned for this task. Following \cite{rappa}, the coherence time measured in seconds is given by $T_{c} = 0.423 / f_{d}$
and the coherence time measured in the number of OFDM symbols, $N_{c}$, can be computed as
\begin{equation}\label{eqn:Nc}
	N_{c} = \frac{\Delta f}{f_{d}}\frac{0.423K}{K+L_{CP}}.
\end{equation}
The effective transmitted power at the UE can be defined as $P_{x}^{\text{eff}}\triangleq\frac{P_{x} }{\eta_{c}}$, where $\eta_{c}$ represents the efficiency factor that takes into account the coherence time and the number of RIS passive elements to be sounded. This efficiency factor is given by 
\begin{equation}\label{eqn:effective_power}
	\quad \eta_{c} \triangleq 1-\frac{T_{r}}{T_{c}} \leq 1-\frac{M}{N_{c}}.
\end{equation}
According to \cite{Ris04}, the time and/or power resources devoted for performing the cascaded channel sounding are penalizing the overall performance of the system.

\subsection{Geometric Wideband Channel Model}
For a more realistic performance evaluation, the two links (BS-RS and RS-UE) are characterized with a geometric wideband model \cite{chan_meas1,chan_meas2}, made up of the superposition of several separate clusters, where each of them has a different value of delay and gain. Note that this channel model is able to account for the spatial correlation considering both the given antenna array response of the BS and RS, as well as the geometrical positions of all clusters/rays. The propagation channel model recommended for 5G \cite{nr-901} is chosen. The power-delay profile follows an exponential distribution whose standard deviation is the DS; the azimuth angles of arrival/departure are modeled by a wrapped Gaussian distribution which is characterized by the Azimuth angular Spread of Arrival and Departure (ASA and ASD); and the zenith angles of arrival/departure are modeled by a Laplacian distribution, also characterized by the Zenith angular Spread of Arrival and Departure (ZSA and ZSD).

A summary of the simulation parameters is provided in Table \ref{tab:simparam}, the location of each network node is given by the Cartesian coordinates $(x,y,z)$ measured in meters, and $f_{c}$ denotes the carrier frequency. The channel propagation model adopted for the simulation results corresponds to the 3GPP factory scenario of size (60m,120m,3m), with the goal to evaluate the 5G performance \cite{nr-901}. Regarding the phase configurations for NCDS, these values are randomly chosen no matter the coherence time.

\begin{table}[!t]
	\centering
	\caption{Simulation Parameters}
	\label{tab:simparam}
	\begin{tabular}{|c|c|c|c|c|c|}
		\hline
		\textbf{BS location}   & (0,0,3) & \boldsymbol{$f_{c}$}     & 3.5 GHz  & \textbf{ASD} & $7^{\rm o}$, $30^{\rm o}$  \\ \hline
		\textbf{RIS location}  & (3,0,3) & \boldsymbol{$\Delta f$}  & 30 KHz   & \textbf{ASA} & $12^{\rm o}$, $50^{\rm o}$  \\ \hline
		\textbf{UE init. location}   & (6,1,1) & \boldsymbol{$K$}           & 1024     & \textbf{ZSD} & $25^{\rm o}$, $130^{\rm o}$ \\ \hline
		\boldsymbol{$L_{\alpha}$}  & -48 dB  & \boldsymbol{$\sigma_{v}^{2}$}  & -94 dBW & \textbf{ZSA} & $30^{\rm o}$, $150^{\rm o}$ \\ \hline
		\boldsymbol{$L_{\beta}$}   & -59 dB  & \boldsymbol{$N$}  & 140 symb.      & \textbf{DS}  & $0.15$ ms   \\ \hline
	\end{tabular}
\end{table}

\subsection{SINR results}

Figure~\ref{fig:simu_sinr_iid} shows the SINR performance as a function of the UE transmit power $P_x$ in dBW of the proposed NCDS for IID Rayleigh channel model. As clearly shown, the performance is better when either the number of antennas at the BS or number of passive elements of the RS are increased. It is also plotted in this figure that the SINR analysis given in (\ref{eqn:diff_snr_iid}), shown with black solid lines, accurately characterizes the RS-empowered system performance.

Figure~\ref{fig:simu_sinr_iid_geo} illustrates the SINR performance comparison as a function of the UE transmit power $P_x$ in dBW of the proposed NCDS between the IID Rayleigh and the geometric wideband channel models, considering $B=2\times 2$ antennas at the BS and different values $M$ for the number of RIS passive elements. As expected, the performance for the IID channel model corresponds to the best case for all simulated $M$ values. On the other hand, for the particular case of geometric wideband channel, the performance depends on the spatial correlation. When the angular positions of the clusters/rays are separated (i.e., high AS), the performance is better compared to the low AS case. 

\subsection{CDS vs NCDS}
The efficiency factor for the CDS, as defined in (\ref{eqn:effective_power}), is evaluated considering the parameters given in Table \ref{tab:simparam}, which are taken from the 5G numerology \cite{nr-901}. The CSI of the cascaded channel is obtained hypothetically assuming that the first $M$ OFDM symbols out of $N_{c}$ (coherence time) are exclusively devoted for reference signal transmission. This is a larger overhead than supported in the 5G standard, but, as explained in Section~\ref{sec:baseline}, it is the minimum that allows a CDS-based RS. In Table \ref{tab:efficiency}, the efficiency factor is numerically evaluated for various values $M$ of the number of the RS passive elements and UE speeds according to (\ref{eqn:effective_power}). As indicated, an RS equipped with large numbers of elements can be only applied in scenarios without or with very low mobility, while an RS with small numbers of elements can be exploited in scenarios with some mobility. It is noted that the exploitation of an RS with large $M$ values using CDS will also have a negative impact on the system complexity.

\eAddFig{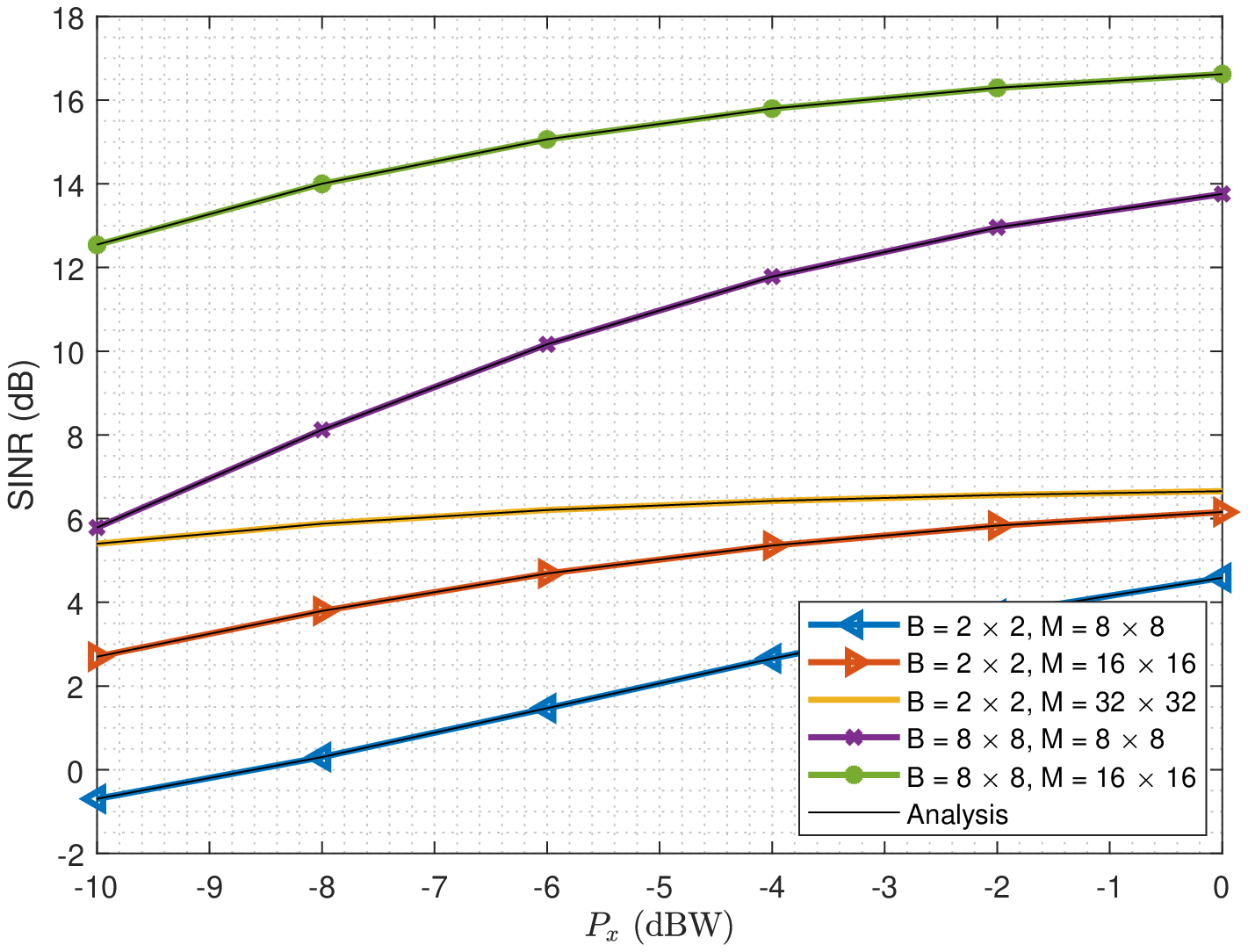}{1}{fig:simu_sinr_iid}{SINR performance in dB of the proposed NCDS for the IID Rayleigh channel model for several values of number of antennas $B$ and passive elements $M$.}
\eAddFig{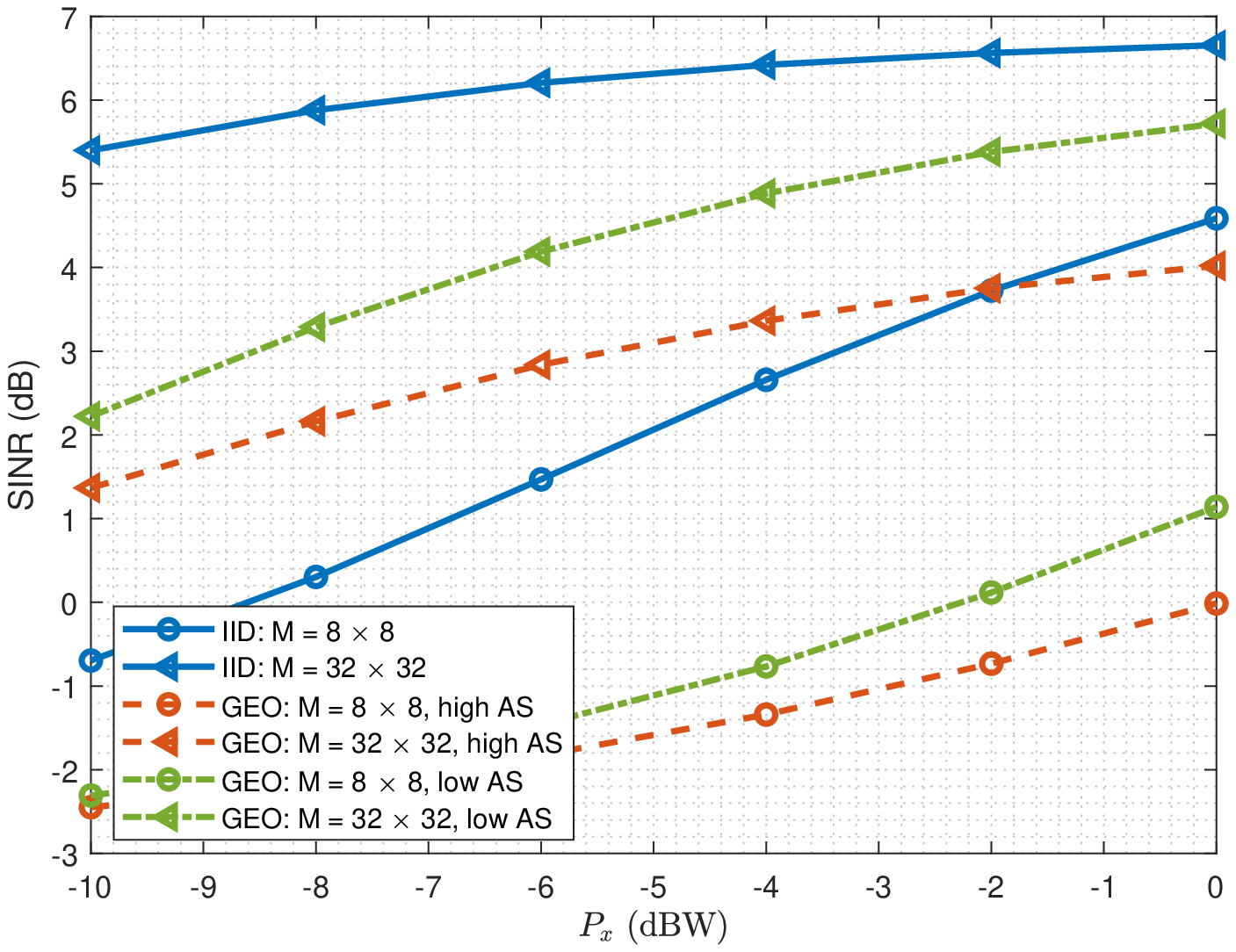}{1}{fig:simu_sinr_iid_geo}{SINR performance in dB of the proposed NCDS for the IID Rayleigh and the geometric wideband channel models for various values $M$ of the RS passive elements, AS, and for $B=2\times 2$ BS antenna elements.}

The performance comparison in terms of SEP between the proposed NCDS and the considered baseline CDS is illustrated in Fig.~\ref{fig:simu_ser_geo_vs} for $4$-DPSK and QPSK modulations, respectively. To provide a fair comparison between the two schemes, we have defined the efficiency factor $\eta=\eta_{c}$ for CDS and $\eta=1$ for NCDS. This highlights that NCDS does not suffer any penalization, unlike CDS. Moreover, it can be deployed at any mobility scenario, exploiting the fact that the differential modulation can be implemented in the frequency dimension, as shown in \cite{Kun2019}. As also concluded from Table~\ref{tab:efficiency}, the proposed NCDS significantly outperforms CDS for large $M$ values for the number of RIS passive elements. Moreover, even for the small $M$ values, the NCDS outperforms CDS, due to the fact that the latter is not able to obtain accurate channel estimates. This happens because the UE's transmit power is in general limited, as also are the amount of resources that can be devoted for CSI estimation while maintaining a reasonable efficiency.

\begin{table}[!t]
	\centering
	\caption{Efficiency Factor for the CDS}
	\label{tab:efficiency}
	\vspace{-3mm}
	\begin{tabular}{|c|c|c|c|c|c|}
		\hline
		& \textbf{3 km/h} & \textbf{10 km/h} & \textbf{20 km/h} & \textbf{30 km/h} & \textbf{40 km/h} \\ \hline
		\boldsymbol{$M$}$\mathbf{=32}$  & 0.9475          & 0.8251           & 0.6484           & 0.4754           & 0.3043           \\ \hline
		\boldsymbol{$M$}$\mathbf{=64}$  & 0.8951          & 0.6503           & 0.2967           & 0                & 0                \\ \hline
		\boldsymbol{$M$}$\mathbf{=128}$ & 0.7902          & 0.3005           & 0                & 0                & 0                \\ \hline
		\boldsymbol{$M$}$\mathbf{=256}$ & 0.5803          & 0                & 0                & 0                & 0                \\ \hline
		\boldsymbol{$M$}$\mathbf{=512}$ & 0.1607          & 0                & 0                & 0                & 0                \\ \hline
	\end{tabular}
\end{table}

\eAddFig{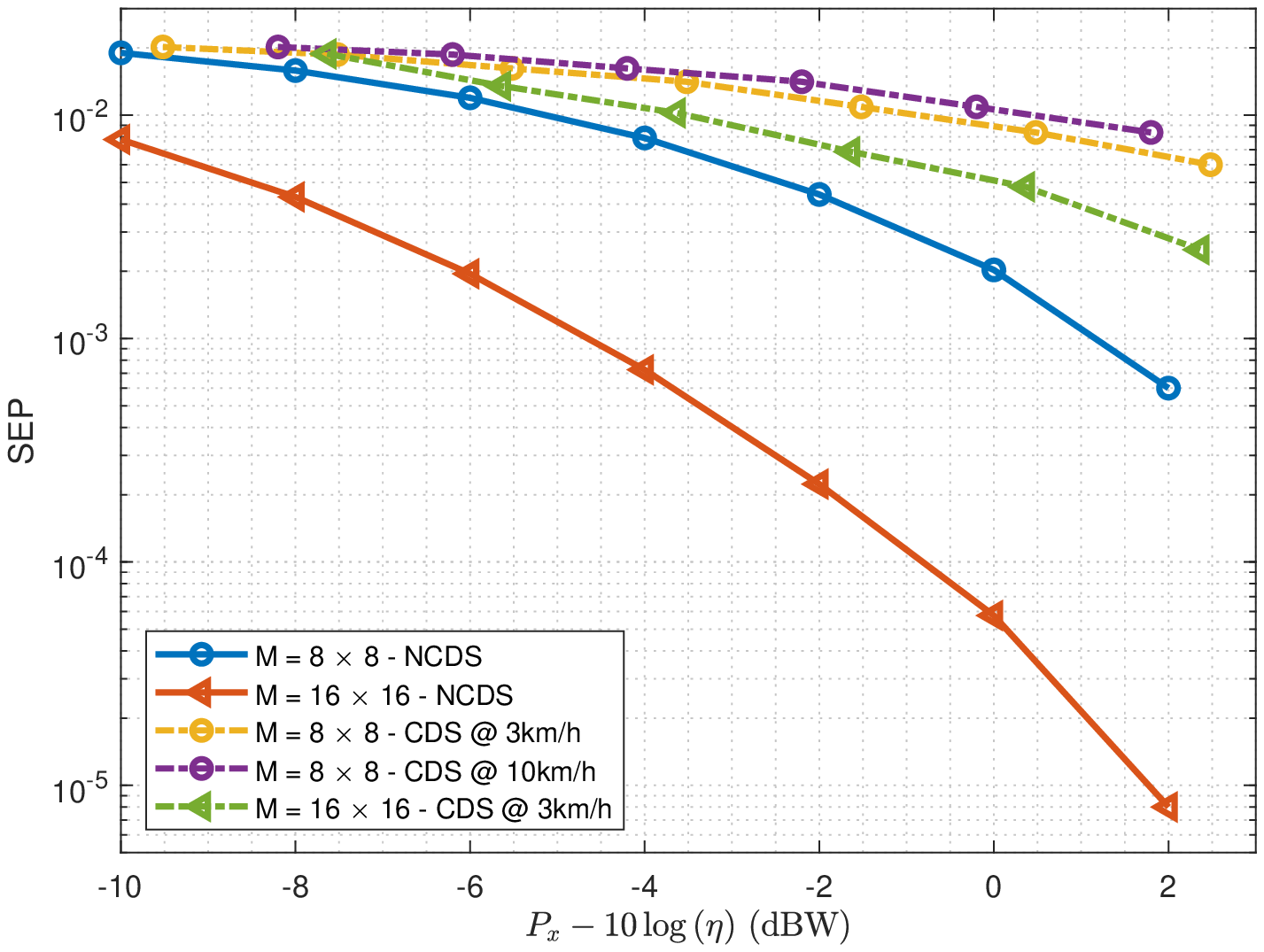}{1}{fig:simu_ser_geo_vs}{SEP performance comparison between the proposed NCDS and the baseline CDS for $4$-DPSK and QPSK, respectively, various numbers $M$ for the RIS elements, $B=2\times 2$ BS antennas, and using a geometric wideband channel model with low AS (ASD=$7^{\rm o}$, ASA=$12^{\rm o}$, ZSD=$25^{\rm o}$, and ZSA=$30^{\rm o}$).}

\section{Conclusion}\label{sec:conclusion}
This paper studied NCDS based on differential decoding as a promising technology for RS-empowered OFDM wireless communications. The proposed scheme is capable of transmitting information avoiding any channel training stage, where neither pilot symbols nor the requirement for solving complex design optimization problems for the BS and RS parameters are needed. The NCDS will enable the advantages of massive numbers of RS passive elements, as well as supporting medium/high mobility and/or low-SNR scenarios. In contrast to CDS, the presented analysis of the SINR revealed that the NCDS performance is not only improved by a large number of BS antennas, but it can be strongly improved by enlarging the number of the RS passive unit elements. 

\section*{Acknowledgment}
This work has been funded by the Spanish National project IRENE-EARTH (PID2020-115323RB-C33 / AEI / 10.13039/501100011033) and European EU H2020 RISE-6G.

\ifCLASSOPTIONcaptionsoff
\newpage
\fi


\end{document}